\documentclass[aps,prb,twocolumn,showpacs,superscriptaddress]{revtex4-2}

\usepackage{graphicx}
\usepackage{amsmath}
\usepackage{multirow}
\usepackage{tabularx}
\usepackage{color}
\usepackage{float}
\usepackage[colorlinks,linkcolor=blue,anchorcolor=blue,citecolor=blue]{hyperref}

\begin{document}

\title{\textit{Ab initio} study on spin fluctuations of itinerant kagome magnet FeSn}

\author{Yi-Fan Zhang}
\affiliation{Center for Neutron Science and Technology, Guangdong Provincial Key Laboratory of Magnetoelectric Physics and Devices, State Key Laboratory of Optoelectronic Materials and Technologies, School of Physics, Sun Yat-Sen University, Guangzhou 510275, China}

\author{Xiao-Sheng Ni}
\affiliation{Center for Neutron Science and Technology, Guangdong Provincial Key Laboratory of Magnetoelectric Physics and Devices, State Key Laboratory of Optoelectronic Materials and Technologies, School of Physics, Sun Yat-Sen University, Guangzhou 510275, China}

\author{Trinanjan Datta}
\affiliation{Department of Chemistry and Physics, Augusta University, 1120 15$^{th}$ Street, Augusta, Georgia 30912, USA}
\affiliation{Kavli Institute for Theoretical Physics, University of California, Santa Barbara, California 93106, USA}

\author{Meng Wang}
\affiliation{Center for Neutron Science and Technology, Guangdong Provincial Key Laboratory of Magnetoelectric Physics and Devices, State Key Laboratory of Optoelectronic Materials and Technologies, School of Physics, Sun Yat-Sen University, Guangzhou 510275, China}

\author{Dao-Xin Yao}

\affiliation{Center for Neutron Science and Technology, Guangdong Provincial Key Laboratory of Magnetoelectric Physics and Devices, State Key Laboratory of Optoelectronic Materials and Technologies, School of Physics, Sun Yat-Sen University, Guangzhou 510275, China}

\author{Kun Cao}
\email{caok7@mail.sysu.edu.cn}
\affiliation{Center for Neutron Science and Technology, Guangdong Provincial Key Laboratory of Magnetoelectric Physics and Devices, State Key Laboratory of Optoelectronic Materials and Technologies, School of Physics, Sun Yat-Sen University, Guangzhou 510275, China}

\begin{abstract}
Kagome antiferromagnetic metal FeSn has become an attracting platform for the exploration of novel electronic states, such as topological Dirac states and the formation of flat bands by localized electrons. Apart from the electronic properties, Dirac magnons and flat magnon bands have also been proposed by applying simplified Heisenberg models to kagome magnetic systems.Inelastic neutron scattering studies on FeSn found well defined magnon dispersions at low energies,but magnons at high energies are strongly dampled, which can not be explained by localized spin models. In this paper, we utilize both linear spin wave theory and time-dependent density functional perturbation theory to investigate spin fluctuations of FeSn. Through the comparison of calculated spin wave spectra and Stoner continuum, we explicitly show that the damping of magnons at high energies are due to the Landau damping, and the appearance of high energy optical-magnon like branches at the M and K point are resulted by relatively low Stoner excitation intensity at those regions. 
      
\end{abstract}

\maketitle

\section{Introduction}\label{introduction}

Kagome lattices have attracted much attention for their ability to host both linearly dispersive bands and dispersionless flat bands. In the pursuit of exotic electronic phenomena, the family of Fe$_x$Sn$_y$ kagome magnets has been of particular interest. Materials in this family are ideal platforms to investigate topological states such as the intrinsic Chern state \cite{yin2020quantum,thouless1982quantized,haldane1988model,xu2015intrinsic}, the fractional quantum Hall effect \cite{tang2011high,sheng2011fractional,neupert2011fractional,sun2011nearly}, high-temperature superconductivity \cite{miyahara2007bcs} and Weyl semimetals \cite{liu2019magnetic}.

As is shown in Fig.~\ref{fig1}(a)(b), FeSn crystallizes in a hexagonal structure (space group P6/mmm). Layers of Fe$_3$Sn can be regarded as kagome lattices of Fe filled with Sn in hexagonal holes, with Fe$_3$Sn layers and Sn layers arranged alternately, forming the crystal structure of FeSn. The corresponding reciprocal unit cell and typical high symmetry {\bf k} points are also illustrated in Fig.~\ref{fig1}(d). As indicated by Mossbauer \cite{ligenza1971mossbauer} and neutron diffraction experiments \cite{yamaguchi1967neutron}, FeSn is antiferromagnetic with the N$\rm \acute{e} $el temperature of 365 K, and the Fe spins are ordered ferromagnetically within the planes, while each successive plane contains spins with opposite directions, as illustrated in Fig.~\ref{fig1}(a),(b). Destructive interference of electronic hopping pathways are proposed to produce nearly localized
electrons, corresponding to flat electron bands in the reciprocal space \cite{han2021evidence,leykam2018artificial,ghimire2020topology}.Similar to the electronic structure, a localized spin model with nearest-neighbor magnetic exchange, $J_1$, generates a flat magnetic band and a Dirac magnon \cite{owerre2017magnonic,mook2014magnon,chisnell2015topological}. This picture is, however, oversimplified in the itinerant kagome magnet FeSn, where interactions between spin wave excitations and the particle-hole continuum of the Stoner excitation could be strong in high-energy area, resulting in spin wave decay, which was first researched by Landau and is currently referred to as Landau damping \cite{landau1946electronic}. The spin fluctuation spectra of FeSn have recently been investigated using inelastic neutron scattering experiment \cite{do2022damped,xie2021spin}. Using low energy incident neutrons, it was observed that the dispersive magnons in the [H, K] plane of the reciprocal space is well defined below 80 meV, while the magnon bandwidth is below 20 meV along the out-of-plane direction. Ref.~\cite{do2022damped} also reported well defined magnons along the whole M-K direction, with the magnon energy reaching about 130 meV at the Dirac K point. Morever, at the M and K points, extended scattering intensity like optical magnons are observed. However, at higher energies, spin waves experience decay and the overall scattering intensity above 80 meV is about 4 orders lower than that of the low energy ones, which was assumed to be mainly caused by Landau damping. Ref.~\cite{xie2021spin} also discussed possible effects of the flat bands on Stoner continuum and magnon excitations.

So far, theoretical studies on the spin fluctuation of FeSn were based on the mapping of itinerant electron systems to effective Heisenberg spin Hamiltonians \cite{do2022damped,xie2021spin,sales2019electronic}, with model parameters fitted to experimental measurements. Corresponding linear spin wave theory (LSWT) calculations then obtained magnon dispersions matching experimental results at low energies. However, they failed to explain the damping of magnons at higher energies due to the localized spin nature of these approaches, which are also sensitive to the determination of their parameters and the choice of discrete spin models. 


In this work, we utilize two methods to investigate spin fluctuations of FeSn, LSWT with model parameters fitted to \textit{ab initio} calculations and  a fully \textit{ab initio} approach based on the analysis of the transverse dynamic magnetic susceptibility within the time-dependent density functional perturbation theory (TD-DFPT )\cite{buczek2011different,savrasov1998linear,rousseau2012efficient}. TD-DFPT is state of the art theoretical tool to study spin fluctuations, computing generalized spin susceptibility directly from \textit{ab initio} electronic structure, thus capable of capturing the interaction between magnons and Stoner continuum. We calculate spin fluctuations spectra along typical high symmetry lines in the Brillouin zone (BZ). At low energies, our LSWT and TD-DFPT results show overall good agreement with experiments, showing a reasonable validity of stand DFT (LDA) in describing magnetic interactions of FeSn. At higher energy region, our TD-DFPT calculations successfully reproduce the shape of the experimental neutron scattering results, including the two optical-magon like branches in the M and the Dirac K points, with a perfect match at the M point, but an overestimated energy at the K point. The correspondence between the calculated spin wave spectra and the Stoner continuum shows explicitly that the cause of the spin wave decay at the K point and other high-energy regions is the Landau damping. The apprearance of the two optical-magnon like branchs is attributed to the existence of two low intensity wedges in the Stoner continuum with coinciding energies to these two branches. 

This paper is organized as follows. In Sec.~\ref{formalism} we outline the framework of our TD-DFPT formalism to calculate the spin fluctuation spectra. In Sec.~\ref{results} we present our calculated results and compare them to that of LSWT, as well as experiments. In Sec.~\ref{discussion} we discuss possible reasons leading to the discrepancies between our theoretical results and experiments. Finally, the conclusions are presented in Sec.~\ref{summary}, together with an outlook on future work.

\section{Formalism}\label{formalism}

In this section, we describe briefly the formalism of TD-DFPT we employed to study the spin fluctuations of FeSn. Unless otherwise stated, Rydberg atomic units are used throughout. A two-step procedure in TD-DFPT is used to calculate the generalized spin susceptibility \cite{buczek2011different,savrasov1998linear,rousseau2012efficient,karlsson2000many,csacsiouglu2010wannier}. Firstly, the Kohn-Sham susceptibility reads
\begin{equation} 
	\begin{aligned}
		\chi_{\mathrm{KS}}^{i j}\left(\mathbf{r}, \mathbf{r}^{\prime}, \omega\right)=& \frac{1}{N_{\mathbf{k}}^{2}} \sum_{n m, \mathbf{k}, \mathbf{q}} \frac{f_{n \mathbf{k}}-f_{m \mathbf{k}+\mathbf{q}}}{\epsilon_{n \mathbf{k}}-\epsilon_{m \mathbf{k}+\mathbf{q}}+\omega} \\
		& \times \vec{\psi}_{n \mathbf{k}}^{\dagger}(\mathbf{r}) \sigma^{i} \vec{\psi}_{m \mathbf{k}+\mathbf{q}}(\mathbf{r}) \vec{\psi}_{m \mathbf{k}+\mathbf{q}}^{\dagger}\left(\mathbf{r}^{\prime}\right) \sigma^{j} \vec{\psi}_{n \mathbf{k}}\left(\mathbf{r}^{\prime}\right)
	\end{aligned} \label{eq:1} 
\end{equation}
where $ N_{\mathbf{k}} $ means the number of $\mathbf{k}$ points used to discretize the first Brillouin zone (we assume a uniform sampling). $\vec{\psi}_{n \mathbf{k}}$ is a Kohn-Sham two-spinor eigenfunction with wave vector $\mathbf{k}$, band index $\textit{n}$, and energy $\epsilon_{n \mathbf{k}}$, and the asterisk indicates complex conjugation. $f_{n \mathbf{k}}$ and $f_{m \mathbf{k}+\mathbf{q}}$ are occupation numbers. The Hartree and exchange-correlation potential are modified by the induced charge and magnetization densities, and the latter two are described by the Kohn-Sham susceptibility, bringing about a self-consistent issue, these densities affect the effective fields and are induced by them at the same time. The second step of this formalism reflects the self-consistency \cite{gross1985local}.
\begin{equation} 
	\begin{aligned}
		&\chi^{i j}\left(\mathbf{r} t, \mathbf{r}^{\prime} t^{\prime}\right) \\
		&=\chi_{\mathrm{KS}}^{i j}\left(\mathbf{r} t, \mathbf{r}^{\prime} t^{\prime}\right)+\sum_{k l} \int d\left(\mathbf{r}_{1} t_{1}\right) d\left(\mathbf{r}_{2} t_{2}\right) \\
		&\quad \times \chi_{\mathrm{KS}}^{i k}\left(\mathbf{r} t, \mathbf{r}_{1} t_{1}\right)\left[f_{x c}^{k l}\left(\mathbf{r}_{1} t_{1}, \mathbf{r}_{2} t_{2}\right)+\frac{2 \delta_{k 0} \delta_{10} \delta\left(t_{1}-t_{2}\right)}{\left|\mathbf{r}_{1}-\mathbf{r}_{2}\right|}\right] \\
		&\quad \times \chi^{l j}\left(\mathbf{r}_{2} t_{2}, \mathbf{r}^{\prime} t^{\prime}\right)
	\end{aligned} \label{eq:2} 
\end{equation}
And the frequency-independent functional derivative of the LSDA exchange-correlation potential is used to approximate the exchange-correlation kernel $f_{x c}^{i j}$ \cite{buczek2011different,gross1985local,von1972local,callaway1975transverse}:
\begin{equation} 
	\begin{aligned}
		f_{x c}^{i j}\left(\mathbf{r} t, \mathbf{r}^{\prime} t^{\prime}\right)=\frac{\delta^{2} E_{x c}}{\delta \rho^{i}(\mathbf{r}) \rho^{j}\left(\mathbf{r}^{\prime}\right)} \delta\left(\mathbf{r}-\mathbf{r}^{\prime}\right) \delta\left(t-t^{\prime}\right)
	\end{aligned} \label{eq:3} 
\end{equation}

The main drawback of this procedure is the requirement of computing unoccupied Kohn-Sham states, to evaluate $\chi^{KS}$ through Eq. (\ref{eq:2}). The convergence with respect to these empty bands is usually very slow. An alternative approach, based on solving the Sternheimer equation, allows us to circumvent this bottleneck by only computing the occupied Kohn-Sham states, and the results are consistent with the first method.

The Sternheimer equation reads
\begin{equation} 
	\begin{aligned}
		\left(\hat{H}-i\frac{\partial}{\partial t} \hat{I}\right)\delta \vec{\psi}_{n \mathbf{k}}(\mathbf{r}, t)=-\left(1-\hat{P}_{\mathrm{occ}}\right) \delta \hat{V}_{\mathrm{scf}}(\mathbf{r},t) \vec{\psi}_{n \mathbf{k}}(\mathbf{r})
	\end{aligned} \label{eq:4} 
\end{equation}
where $\hat{H}$ denotes the unperturbed Kohn-Sham Hamiltonian, corresponding to the term in square brackets, $\delta \vec{\psi}_{n \mathbf{k}}$ is the first-order change of the spinor wave function, and $\delta\hat{V}_{\mathrm{scf}}$ is the first-order variation of the Kohn-Sham potential. The operator $\hat{P}_{\mathrm{occ}}$ is the projector on the manifold of unoccupied Kohn-Sham states.  Under this scheme, a self-consistent ground state calculation is first performed , obtaining ground state electron density and Kohn-Sham wavefunctions.  An external plane-wave magnetic field is then applied to the right hand side of Eq.~\ref{eq:4} which is then solved self-consistently to obtain the first-order response of spin density,  therefore the generalized spin susceptibility. Our implementation is based on the linear-response modules of the QUANTUM ESPRESSO materials simulation suite \cite{giannozzi2009quantum}.More details of this formalism are shown in Ref.~\cite{cao2018ab}.

Our \textit{ab initio} calculations were performed with the Quantum Espresso suite \cite{giannozzi2009quantum}. We used the local density approximation for the exchange and correlation functional \cite{perdew1981self} and corresponding ultrasoft pseudopotentials from the repository 'PSlibrary 0.3.1' \cite{dal2014pseudopotentials}. We employed a plane-waves kinetic energy cutoff of 40 Ry, and the Gaussian smearing technique was used to deal with the Brillouin zone (BZ) integration in the presence of a Fermi surface, with a width of 30 mRy. In order to accommodate the layered antiferromagnetic order as observed experimentally, we used a lattice comprising doubled unit cell along the c direction with experimental lattice parameters $\textit{a}$ = 5.305 $\AA$  and $\textit{c}$ = 8.928 $\AA$. A 12 × 12 × 6 grid of $\textbf{k}$ points was used for both the ground state calculation and the solution of the Sternheimer equation. Our LSWT calculations are performed with SpinW software package \cite{toth2015linear}.

\section{Results}\label{results}

\subsection{Linear spin wave theory}\label{LSWT}
	
\begin{figure}[t]
	\centering
	\includegraphics[scale=0.35]{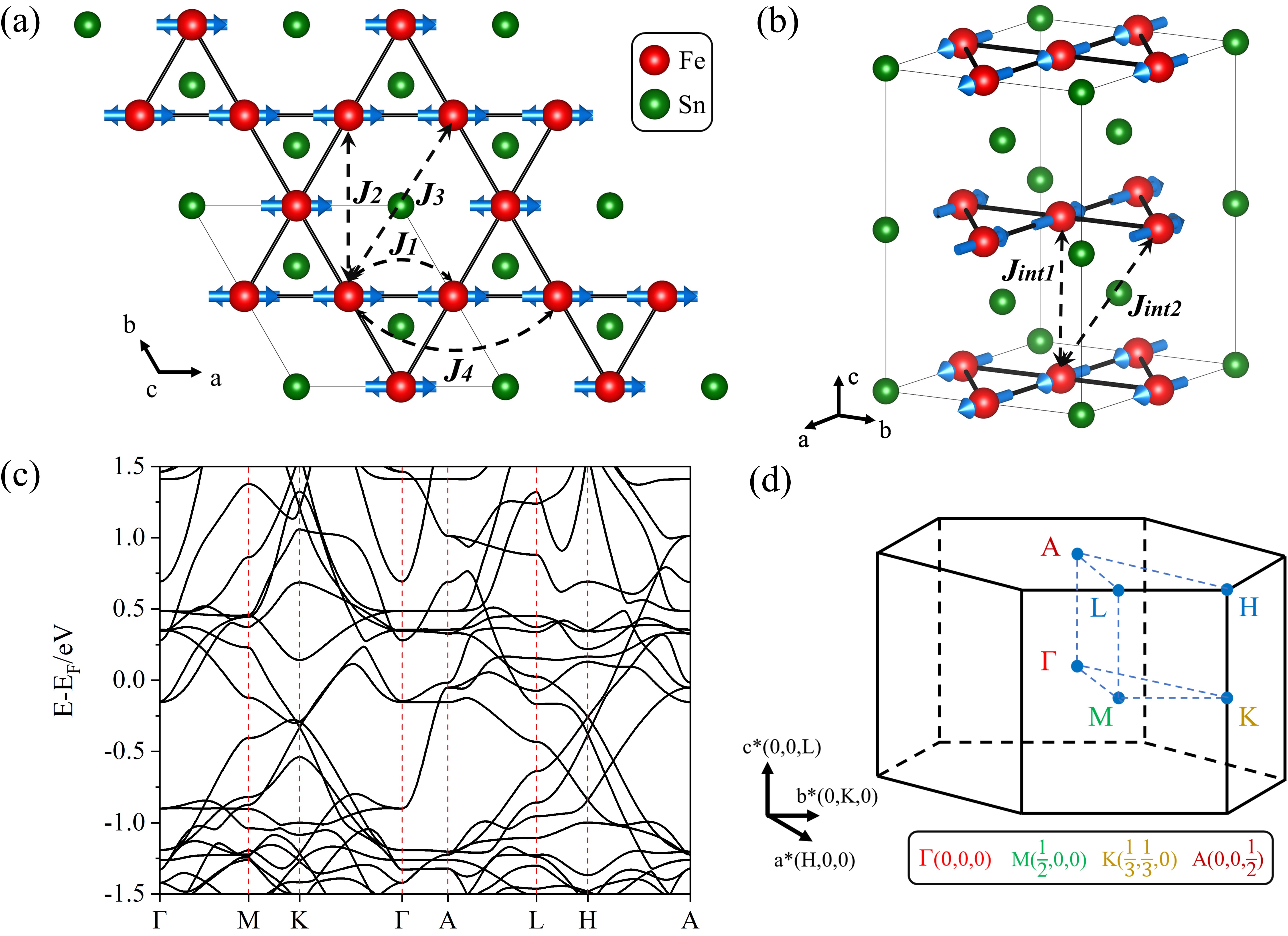}
	\caption{(a),(b) Crystal and magnetic structures of FeSn. Red spheres represent Fe atoms and Green spheres represent Sn atoms. Arrows denote magnetic moments of Fe atoms. (c) Calculated electronic band structures of FeSn.  (d) 3D Brillouin zone of FeSn with the selected $\Gamma$-M-K-$\Gamma$-A-L-H-A high-symmetry directions.}
	\label{fig1} 
\end{figure}

Our \textit{ab initio} calculations produce a magnetic moment of 1.89 $\mu_B$/Fe, in good agreement with experimental values of 1.85 $\mu_B$/Fe \cite{sales2019electronic}. The calculated band structure is shown in Fig.~\ref{fig1} (c), with clearly visible Dirac K point and flat bands about 0.5 eV above the Fermi energy, consistent with electronic structure calculated in Ref.~\cite{lin2020dirac}. We then map the magnetic interactions of itinerant FeSn to the following Heisenberg model, 
\begin{equation} 
	\begin{aligned}
		\textit{H} = J_{ij}\sum_{i<j}\textbf{S}_i\ \cdot\ \textbf{S}_j - D_{z}\sum_{i}\left(\textbf{S}_{i}^{z}\right)^{2}
	\end{aligned}
\end{equation}	
where \textbf{S}$_i$ corresponds to local spin of Fe ions. As is shown in Fig.~\ref{fig1}(a) and Fig.~\ref{fig1}(b), we consider four types of intra-layer exchange interactions $J_{1}$-$J_{4}$ and two types of inter-layer exchange interactions $J_{int1}$ and $J_{int2}$. $J_{1}$ and $J_{2}$ denote nearest neighbour and next nearest neighbour interactions respectively. $J_{3}$ and $J_{4}$ represent two inequivalent third nearest neighbour interactions. $D_{z}$ corresponds to the single ion anisotropy. Our model is consistent with that of Ref.~\cite{do2022damped}, taking into account additional inter-layer interactions $J_{int2}$ and intra-layer interactions $J_{3,4}$, in comparison to Ref.~\cite{xie2021spin} and Ref.~\cite{sales2019electronic}. The $J$'s and $D$ are obtained by fitting to energies from constrained \textit{ab initio} calculations with different noncollinear spin configurations.
\begin{table}
	\caption{\label{tab:table1} Heisenberg exchange
		couplings from the linear spin wave theory.}
	\begin{ruledtabular}
		\begin{tabular}{cccccccc}
			&$J_{1}$&$J_{int1}$&$J_{2}$&$J_{int2}$&$J_{3}$&$J_{4}$&$D_{z}$\\ \hline
			J$_{ij}$/$meV$&-53.26&35.96&3.46&-6.83&1.49&1.89&0.001\\
			Distance/$\AA$&2.652&4.464&4.594&5.193&5.305&5.305&-\\	
		\end{tabular}
	\end{ruledtabular}
\end{table}

The calculated value of each $J$ is listed in Table \ref {tab:table1}. We can see that the nearest neighbour ferromagnetic $J_{1}$ dominates intralayer interactions, as a result, the intralayer magnetic structure is ferromagnetic, although $J_{2}$, $J_{3}$ and $J_{4}$ are antiferromagnetic. It is worth noting that the inter-layer $ J_{int1}$ is strongly antiferromagetic with its amplitude in the same order of $J_{1}$, which is unusual given its relatively long bond length. This might be a manifestation of itinerant character of interlayer exchange interactions of FeSn. However, the effect of antiferromagnetic $J_{int1}$ is partly cancelled by the ferromagnetic $J_{int2}$, whose coordination number is four times larger than that of $J_{int1}$, leading to a much weaker overall antiferromagnetic interlayer interaction, similar to other theoretical models with only one interlayer interaction \cite{xie2021spin}. To further verify our calculated parameters, we also performed classical monte carlo simulations with this set of parameters, obtaining a correct magnetic phase diagram and a T$_{N}$ about 392 K, which is in good agreement with the experimental T$_{N}$ =365 K \cite{sales2019electronic}.

The calculated spin wave spectrum along high symmetry  $\Gamma$-M-K-$\Gamma$-A-L-H-A directions in the BZ are shown as green lines in Fig.~\ref{fig2}(a), with the position of each high symmetry point shown in Fig.~\ref{fig1}(d). We can see that along the $\Gamma$-M direction, the excitation energy reaches about 75 meV at the M point, and further reaches 120 meV along the M-K direction around the K point, where a linear crossing  indicating the existence of Dirac magnons. Unlike the in-plane magnon dispersion, the out-of-plane magnon dispersion along $\Gamma$-A direction can only reach around 20 meV at the A point, indicating a much weaker interlayer coupling, consistent with our analysis on the calculated model parameters. By comparing our LSWT results with experimental excitation energies in Fig.~\ref{fig2}(a), we can see an overall good agreement in the low-energy region below 75 meV, with the position of experimental excitations falling slightly above the calculated dispersion along $\Gamma$-M and $\Gamma$-K directions, but gaining a better match along the rest of the selected high symmetry directions. This agreement, in general, shows that LDA can describe magnetic interactions in FeSn reasonably well. However, the LSWT does not capture the experimentally observed optical-magnon like branch from about 80 to 160 meV at the M point and above 150 meV at the Dirac K point. Morever, the experimental spectra show decay of magnons at energies above 120 meV, and the higher-energy spectral weight becomes indiscernible from the background scattering. This observed decay of magnons is proposed to be caused by the Landau damping, which can not be properly taken into account in the LSWT. 

\subsection{Time-dependent density functional perturbation theory}\label{TDDFPT}

\begin{figure*}[!htb]
	\centering
	\includegraphics[scale=0.8]{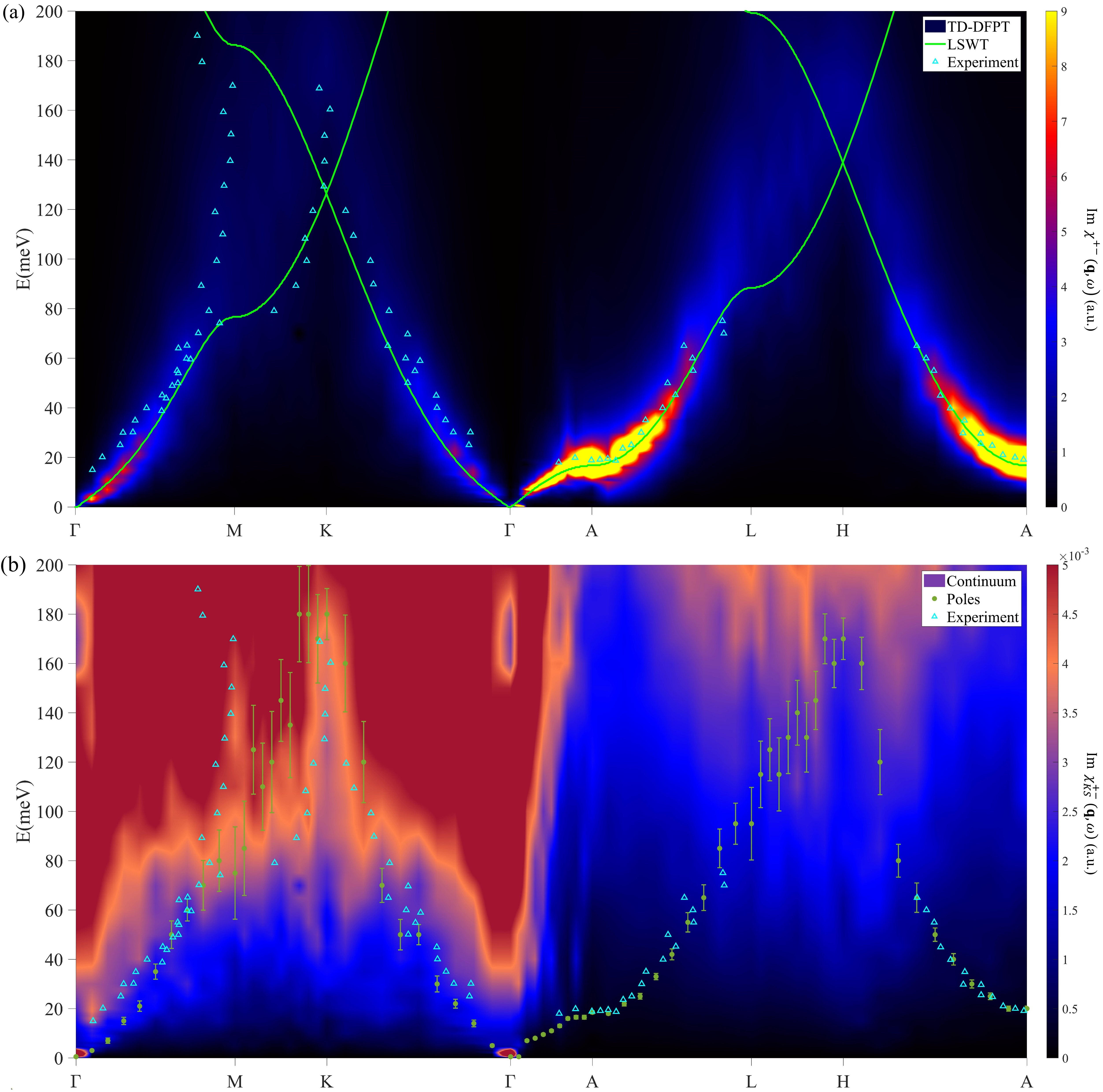}
	\caption{(a) Spin fluctuation spectra of FeSn along the high-symmetry directions below 200 meV. The color map represents the calculated  Im$\chi_{K S}(\mathbf{q}, \omega)$. The results of LSWT are shown as green lines and the experimental data as triangles\cite{do2022damped}. (b) Calculated Stoner excitation spetra Im$\chi_{K S}^{+-}(\mathbf{q}, \omega)$. The green dots correspond to the poles of Im$\chi^{+-}$, and the vertical error bars correspond to the width at the half maximum (FWHM) of the excitation peaks, which are reduced to one fifth of the original values to fit into the figure.}
	\label{fig2} 
\end{figure*}

As is described in Sec. \ref{introduction}, the standard image of spin wave in itinerant ferromagnets can be distilled to: the energy of the magnon increases as the momentum rises and its lifetime keeps infinite until the magnon band comes into contact with the Stoner continuum, the peak linewidths then gradually broaden and the spin wave cannot be viewed as a well-defined quasi-particle excitation. In the spirit of related work in the area of linear response theory, the whole information of the spin-wave spectra and the Stoner continuum can be obtained by calculating the transverse spin susceptibility $\chi^{+-}(\mathbf{q}, \omega)$ and the corresponding non-interacting Kohn-Sham susceptibility $\chi_{KS}^{+-}(\mathbf{q}, \omega)$, where magnon excitation energies correspond to the poles of Im$\chi^{+-}(\mathbf{q}, \omega)$, with the full width at the half maximum (FWHM) of each peak provides the inverse lifetime of each excitation and the nonzero region in the $(\mathbf{q}, \omega)$ plane of Im$\chi_{K S}(\mathbf{q}, \omega)$ forms the Stoner excitation continuum. 



\begin{figure}
	\centering
	\includegraphics[scale=0.48]{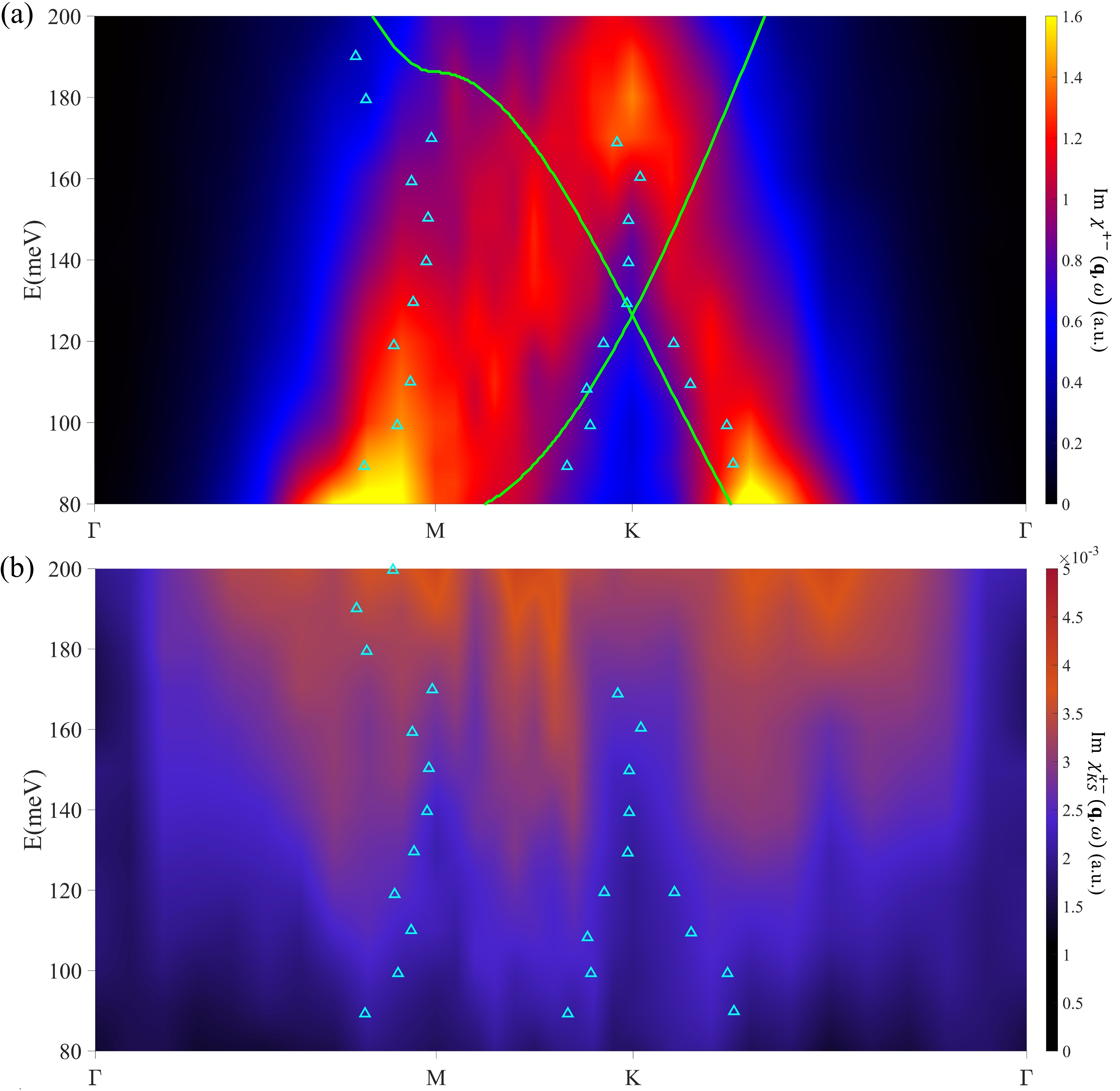}
	\caption{(a) The color map represents Im$\chi^{+-}(\mathbf{q}, \omega)$ above 80 meV. The green lines represent the results of LSWT and the triangles represent experimental data. (b) Calculated Stoner excitation spectra Im$\chi_{K S}^{+-}(\mathbf{q}, \omega)$, averaged over L=0, 0.25, 0.5, 0.75, 1, 1.5, 2 to enhance statistics.}
	\label{fig3} 
\end{figure}

Using TD-DFPT described in Section. \ref{formalism}, we calculated Im$\chi$$^{+-}$($\textbf{q}$,$\omega$) for wave vectors $\textbf{q}$ along the $\Gamma$-M-K-$\Gamma$-A-L-H-A high-symmetry lines, with energy ranging from 0 to 200 meV. A same k point mesh is used for both the ground state calculations and the solution of the Sternheimer equation, which is required in our formulism to avoid the presence of long-wavelength magnons with finite excitation energies, the so-called gap error which is caused by spurious symmetry breaking \cite{lounis2011theory}. Although our calculations satisfy the Goldstone theorem, the excitation energy shows a tiny gap $\sim$ 0.5 meV due to numerical error. This nevertheless does not affect our conclusions.

The color map of Fig.~\ref{fig2}(a) shows calculated Im$\chi^{+-}(\mathbf{q}, \omega)$ in the ($\textbf{q}$,$\omega$) plane. Well defined spin wave dispersion can be seen below around 80 meV along the whole sampled high symmetry directions. Along the $\Gamma$-M direction, the TD-DFPT results are in good agreement with that from LSWT, thus also sit slightly below the experimental values. While in the $\Gamma$-K direction, the TD-DFPT results are overall above LSWT results and therefore, match experimental values very well. Typical for itinerant magnetic systems, the TD-DFPT calculated magnons experience enhanced damping with the increase of excitation energies, with no well defined magnons identifiable in high energy regions, in consistence with experimental observations. More remarkably, TD-DFPT reproduces the shape of neutron scattering results above 80 meV along the $\Gamma$-M-K-$\Gamma$ direction, including the optical-magnon like branches at the M and K point(Fig.~\ref{fig2}(a)). To show a more accurate comparison with experiments in Ref.~\cite{do2022damped}, we replot $\times$Im$\chi^{+-}(\mathbf{q},\omega)$) at the high energy region along the $\Gamma$-M-K-$\Gamma$ direction in Fig.~\ref{fig3}(a). It can be seen that the calculated excitation energies at the M point agree very well with experiments. While at the K point, the Dirac magnon and its upper band are reproduced, but with energies overestimated by about 30 meV. On the other hand, the TD-DFPT spectra along the A-L-H-A direction resemble that along the $\Gamma$-M-K-$\Gamma$, but exhibiting a shift along the energy axis and much stronger intensity.  The resulting comparison between these two high symmetry directions is not unexpected since the A-L-H-A is obtained by shifting the $\Gamma$-M-K-$\Gamma$  by (0, 0, $\frac{1}{2}$) (see Fig.~\ref{fig1}(d)). 

To explore the effect of Landau damping, we calculate Stoner excitation spectra along the same set of k points, as shown in Fig.~\ref{fig2}(b). Along the $\Gamma$-M-K-$\Gamma$, it clearly shows that the magnon spectra overlap with higher intensity Stoner excitations with the increased excitation energies, especially above 80 meV, therefore result in heavy attenuation in this region. Interestingly, even at high energies, there are two wedges of relatively weaker regions in the Stoner continuum, coinciding with the optical-magnon like branches at the M and K points, which well explain the apprearance of these two branches. In addition, the overall intensity of the Stoner continuum along the $\Gamma$-A-L-H-A direction is much weaker than that in the $\Gamma$-M-K-$\Gamma$ direction, leading to the opposite comparison of the corresponding magnon spectra. To enhance statistical sampling, Ref.~\cite{do2022damped} integrated the high energy neutron scattering data on L over [-4, 4], to show more clearly the excitations around the M and K points. Similarly, we also calculate Stoner continuum with different L values for each [H, K] point. Compared to results with L=0, an average over all sampled L values shows an overall lower Stoner excitation intensity, with the two wedges at the M and K points becoming more significant, therefore could lead to more visible uppper bands at these two points(see Fig.~\ref{fig3}(c)).

\section{discussion}\label{discussion}
Recently, a spin-polarized flat electronic band has been identified in the antiferromagnetic FeSn at an energy 230 ± 50 meV below the Fermi level by angle-resolved photoemission spectroscopy (ARPES) experiments \cite{kang2020dirac}. DFT calculations also suggested the existence of a narrow electronic band with opposite spin lying about 0.5 eV above the Fermi level. The spin-flip electronic transition between these two bands could lead to a narrow Stoner excitation region, which may have noticable impact on magnon excitations, as suggested in Ref.~\cite{xie2021spin}. However, such a narrow Stoner continuum would have excitation energies around 730 meV, which is much higher than typical magnon energies. Morever, at such high energies, Stoner excitations from other transition channels could mix heavily with the narrow band and render the flatness undistinguishable. Unfortunately, standard DFT calculations(LDA, PBE) can not reproduce the correct position of the flat band below the Fermi level, see Fig.~\ref{fig1}(c). and Ref.~\cite{kang2020dirac}, which inevitably lead to inaccurate description of Stoner continuum, even at relatively lower energies(around 200 meV), therefore result in discranpencies between theory and experimental neutron scattering results. Such observations can also be found for simple transition metals, such as Fe and Co, where low energy neutron scattering data can be well reproduced by standard TD-DFPT calculations, but theories deviate from experiments gradually after that Stoner excitations kick in to affect magnon excitations \cite{buczek2011different,savrasov1998linear,rousseau2012efficient,cao2018ab}. Therefore, more elaborate methods, properly taking into account electron correlation, such as DFT+DMFT, may be explored in the future to investigate spin fluctuations in kagome metals. \cite{mcnally2015hund,song2020photoemission,yin2014spin,stadler2015dynamical,yin2011kinetic}

\section{Summary}\label{summary}    

In summary, spin fluctuations of Kagome metal FeSn are investigated by \textit{ab initio} based LSWT and  a fully \textit{ab initio} approach based on the analysis of the transverse dynamic magnetic susceptibility within the time-dependent density functional perturbation theory. The calculated spin fluctuations spectra along typical high symmetry lines in the BZ agrees well with experiments at low energies.  At higher energy region, the TD-DFPT calculations successfully reproduce the shape of the experimental neutron scattering results, including the two optical-magon like branches at the M point and the Dirac K point, with perfectly matched energies at the M point, but an overestimation at the K point. The correspondence between the calculated spin wave spectra and the Stoner continuum shows explicitly that the cause of the spin wave decay at the K point and other high-energy regions is the Landau damping. The appearance of the two optical-magnon like branchs is attributed to the existence of two low intensity wedges in the Stoner continuum with coinciding energy to these two branches. The discrepancies between theory and experiments imply the deficiency of standard DFT to capture the full electron correlation effect in FeSn, therefore more advanced methods, such as DFT+DMFT, is desirable to investigate spin fluctuations of kagome metal systems in the future.

\section{acknowledgments}

Work at Sun Yat-Sen University was supported by the Guangdong Basic and Applied Basic Research Foundation (Grants No. 2022A1515011618, No. 2021B1515120015, No. 2019A1515011337), and the National Natural Science Foundation of China (Grants No. 12174454, No. 11904414, No. 92165204, No. 11974432), and the National Key Research and Development Program of China (Grants No. 2019YFA0705702, 2018YFA0306001, 2017YFA0206203), Shenzhen International Quantum Academy (Grant No. SIQA202102), Leading Talent Program of Guangdong Special Projects (201626003). TD acknowledges the hospitality of KITP at UC-Santa Barbara and support by the National Science Foundation under Grant No. NSF PHY-1748958. 


%

\end{document}